\documentclass[twocolumn]{aastex631}
\usepackage{bm}
\everymath{\displaystyle}

\shorttitle{Particle acceleration by relativistic shocks}
\shortauthors{Morikawa et al.}
\usepackage{natbib}

\begin{document}

\title{Particle acceleration and magnetic field amplification by relativistic shocks in inhomogeneous media}
\author[0009-0006-2453-9563]{Kanji Morikawa}
\author[0009-0006-2453-9563]{Yutaka Ohira}
\affiliation{Department of Earth and Planetary Science, The University of Tokyo, 
7-3-1 Hongo, Bunkyo-ku, Tokyo 113-0033, Japan}
\email{kanji.m1029@eps.s.u-tokyo.ac.jp}


\author[0000-0002-0040-8968]{Takumi Ohmura}
\affiliation{Institute for Cosmic Ray Research, The University of Tokyo, 
5-1-5 Kashiwanoha, Kashiwa, Chiba 277-8582, Japan}
\affiliation{Division of Science, National Astronomical Observatory of Japan, 2-21-1 Osawa, Mitaka, Tokyo 181-8588, Japan}

\begin{abstract}
Particle acceleration and magnetic field amplification in relativistic shocks propagating in inhomogeneous media are investigated by three-dimensional magnetohydrodynamical (MHD) simulations and test-particle simulations.
The MHD simulations show that the interaction between the relativistic shock and dense clumps amplifies the downstream magnetic field to the value expected from observations of the gamma-ray burst.
The test-particle simulations in the electromagnetic field given by the MHD simulation show that particles are accelerated by the downstream turbulence and the relativistic shock.
We provide the injection energy to the shock acceleration in this system.
If the amplitude of upstream density fluctuations is sufficiently large, low-energy particles are initially accelerated to the injection energy by the downstream turbulence and then rapidly accelerated to higher energies by the relativistic shock. 
Therefore, the density fluctuation significantly affects  particle acceleration in the relativistic shock.
\end{abstract}

\section{Introduction}
\label{sec:1}
High-energy charged particles called cosmic rays (CRs) can be mainly separated into two categories, galactic and extra-galactic CRs.
The most promising acceleration mechanism is first-order Fermi acceleration (diffusive shock acceleration) \citep{Axford1977, Krymskii1977, Blandford1978, Bell1978}.
In this process, particles are accelerated by a shock, crossing the shock front many times.
The galactic CRs are thought to be accelerated by this mechanism in supernova remnants in our galaxy. 
On the other hand, the origin of extra-galactic CRs has long been a mystery. 
Naively, the extra-galactic CRs are expected to be accelerated by shocks in the same way as the galactic CRs. 
Since objects that can potentially produce the extra-galactic CRs mostly have relativistic outflows \citep{Hillas1984,Alves2019}, relativistic shock waves have been intensively studied as an accelerator of the extra-galactic CR \citep{Kirk1987,Begelman1990,Gallant1999,Achterberg2001,Lemoine2003,Keshet2005,Lemoine2006,Lemoine+2006,Niemiec2006,Sironi2013,Kirk2022,Huang2023}.

The relativistic shock acceleration needs strong downstream magnetic turbulence to return particles to the upstream region.
For weakly magnetized relativistic shocks, the Weibel instability generates the magnetic turbulence, so that the shock acceleration works \citep{Spitkovsky2008}.
However, the Weibel-mediated shock cannot accelerate CRs with $10^{20}~{\rm eV}$ within the dynamical time scale of astrophysical relativistic shocks because the length scale of the magnetic turbulence is much smaller than the gyroradius of accelerated particles \citep{Sironi2013}.
Particles can be rapidly accelerated by the relativistic shock if the shock generates magnetic turbulence with a larger length scale \citep{Lemoine2006,Kirk2022,Huang2023}. 
In addition, the larger magnetic turbulence in the relativistic shock is required to explain bright prompt or afterglow emissions from gamma-ray bursts (GRBs) \citep{Gruzinov1999, Huang2022}. 
Nevertheless, the generation mechanism of the magnetic turbulence is still an open issue \citep{Tomita2016, Tomita2019}.

In this work, we demonstrate, by solving the generation of magnetic turbulence, that the large-scale magnetic turbulence is generated in the relativistic shock, and then particles are rapidly accelerated in the relativistic shock.
We consider a relativistic shock propagating in an inhomogeneous medium. 
For non-relativistic or mildly relativistic shocks, it is shown that the interaction between the shock and a shock-upstream denser region can drive the strong magnetic turbulence in the shock-downstream region \citep{Giacalone2007, Inoue2011, Mizuno2014, Fraschetti2013, Ohira2016, Ohira2016b, Slavin2017, Romansky2020, Hu2022, Demidem2023, Bresci2023, Fulat2023}.
Although there are few studies about a generation of downstream magnetic turbulence in the case of the relativistic shock \citep{Sironi2007,Goodman2008,Tomita2022}, 
none of them demonstrate the generation of the magnetic field turbulence that can accelerate CRs rapidly and explain emissions from GRBs. 
First of all, we perform three-dimensional relativistic magnetohydrodynamical (MHD) simulations to obtain the magnetic field structure in the downstream region of a relativistic shock propagating to an inhomogeneous medium.
Then, we perform test-particle simulations, showing that particles are rapidly accelerated in the magnetic field obtained in the relativistic MHD simulations.
Hereafter, we use the notation $X_{\rm ,U}, X_{\rm ,S},$ and $X_{\rm ,D}$ to indicate the quantity, $X$, measured in the upstream, shock, and downstream rest frames, respectively.

\section{Numerical simulations}
\label{sec:2}
\subsection{Relativistic MHD simulations}
\label{sec:2_1}
In order for particle acceleration to work in the relativistic shock, particles have to cross magnetic field lines. 
To avoid the unphysical suppression of the particle diffusion perpendicular to the magnetic field line, we need a magnetic field structure in the realistic three-dimensional space for the test-particle simulation \citep{Giacalone1994, Jokipii1993}. 
We used the special relativistic three-dimensional MHD simulation code, SRCANS+ \citep{Matsumoto2019}, where the approximate Riemann solver is the HLL method, the MUSCL method in 2nd order is adopted for the spatial higher-order method, the limiter is the van Leer limiter, and the error of $\boldsymbol{\nabla} \cdot \boldsymbol{B}$ is removed by using the 9-wave method \citep{Dedner2002, Mignone2010}. 
We adopted the equation of state proposed in \citet{Mignone2007} for primitive recovery.
The simulation box size is $(L_x/\Delta x, L_y/\Delta y, L_z/\Delta z)=(8000, 200, 200) $, where $L_s$ and $\Delta s= 0.1c\Delta t_{\rm MHD}$ are the box size and grid size in the $s$ directions ($s=x,y,z$) and $\Delta t_{\rm MHD}$ is the time step. 
The $x$-direction is set to be perpendicular to the mean shock front.
The boundary condition in the $x$-direction is the open boundary and the periodic boundary condition is imposed for the $y$- and $z$-directions.
We set an initial condition so that the shock front averaged over the $y$-$z$ plane appears stationary, that is, the simulation frame is the mean shock rest frame. \par

In the simulation frame, the upstream bulk Lorentz factor is $\Gamma_{\rm up, S}=5.2$ and the upstream density structure is given as follows,
\begin{eqnarray}
\rho_{\rm up, S}(x,y,z) = \rho_{\rm 0, S} + \sum_{i} \delta \rho_{i{\rm ,S}}(x,y,z),
\label{eq:density}
\end{eqnarray}
where $\rho_{\rm 0, S}$ is the constant density and the individual clumps are represented by
\begin{eqnarray}
\delta \rho_{i{\rm , S}}=\left\{
\begin{array}{ll}
~0 & (d > 2L_{\rm cl}) \\
~ \left(\rho_{\rm cl, S} - \rho_{\rm 0, S}\right)\left\lbrace1 + \cos\left(\frac{\pi d}{2L_{\rm cl}} \right)\right\rbrace  & ( d \leqq 2L_{\rm cl})~,
\end{array}
\right.\\
d \equiv \sqrt{(x-x_{{\rm c}i})^2\Gamma_{\rm up, S}^2+(y-y_{{\rm c}i})^2+(z-z_{{\rm c}i})^2},~~~~~&
\label{eq:density distribution}
\end{eqnarray}
where $x_{{\rm c}i},~y_{{\rm c}i}$,~$z_{{\rm c}i}$, $d$, and $L_{\rm cl}=0.1L_y$ are the center of the individual clump, distance from the center, and the half-width of the clumps, respectively. $\rho_{\rm cl,S}$ is the density at the half width. 
We randomly put the clumps in the upstream rest frame. Then, the Lorentz transformation was performed to obtain the center of each clump in the simulation frame.
The shape of the clumps is a sphere in the upstream rest frame. 
The volume filling factor of the density clumps is set to $N_{\rm cl} V_{\rm cl, S}/L_x L_y L_z = 0.2512$ in this work, where $N_{\rm cl}$ and $V_{\rm cl, S}= 4\pi L_{\rm cl}^3/3\Gamma_{\rm up, S}$ are the number of clumps in the simulation box and the volume of each clump, respectively.
Initially, the magnetic field in the upstream rest frame and gas pressure are assumed to be uniform, $B_{\rm up, U}$ and $p_{\rm up}$, and the magnetic field is pointing in the $y$-direction.  
The upstream magnetization parameter and plasma beta are set to $\sigma = B_{\rm up, U}^2/4\pi \rho_{\rm 0, U} c^2=3.7\times 10^{-6}$ and
$\beta = 8\pi p_{\rm up}/B_{\rm up,U}^2=5.4\times10^3$, where $c$ is the speed of light.
In this work, we perform two MHD simulations for $\rho_{\rm cl,S}/\rho_{\rm 0,S}=2$ and $10$.


\subsection{Test-particle simulations}
Since the shock front is almost stationary in our simulation frame and the downstream turbulent velocity is about $c\sim 0.1$, we use the snapshot data of the electromagnetic field obtained by the MHD simulations at $t=3000L_{\rm cl}/c$ to solve the equation of motion for charged particles. This stationary velocity field $\boldsymbol{u}_{\rm ,S}$ can accelerate particles through the electric field, $\boldsymbol{E}_{\rm ,S} = - ( \boldsymbol{u}_{\rm ,S} / c ) \times \boldsymbol{B}_{\rm ,S}$. In order to solve the equation of motion for the charged particles, we adopt the Buneman-Boris method \citep{Birdsall2018} with the time step of $\Delta t_{\rm TP} = 0.02\pi\Delta x/c$. 
$10^6$ simulation particles are uniformly injected in the upstream region. 
The initial momentum distribution is isotropic with the Lorentz factor of $\gamma_{\rm p, U}=10$ in the upstream rest frame.
We follow particle orbits until the particles reach the downstream boundary at $x=300~L_{\rm cl}$. 
Because the shock front is located at $x\sim 240~L_{\rm cl}$ in our MHD simulation, 
the downstream size in the $x$-direction is $L_{\rm down} \sim 60~L_{\rm cl}$.

To understand the effect of clump size on particle acceleration, 
we perform two test-particle simulations ($R_{\rm up, U}/L_{\rm cl}=26$ and $0.26$) for each of the two MHD simulations, 
where $R_{\rm up, U}$ is the initial gyroradius in the upstream region in the upstream rest frame. 

\section{Results}
\label{sec:3}
\begin{figure}[tbp]
    \centering
    \includegraphics[width=\linewidth]{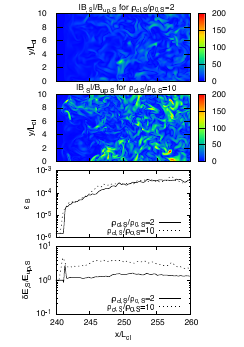}
   \caption{
   MHD simulation results at $t=3000~L_{\rm cl}/c$.
   The top two panels show the two-dimensional distribution of the magnetic field strength in the $z=0$ plane. 
   The third and last panels show one-dimensional distributions of magnetic energy density and electric field strength associated with the downstream turbulence, respectively. 
   The bottom two panels show quantities averaged over the $y$-$z$ plane.}
    \label{fig:result_mhd_b_field}
\end{figure}
Figure~\ref{fig:result_mhd_b_field} shows the MHD simulation results at $t= 3000~L_{\rm cl}/c$.
The top two panels are the two-dimensional distribution of magnetic field strength in the $z=0$ plane for $\rho_{\rm cl,S}/\rho_{\rm 0,S}=2$ and $10$, 
and the bottom two panels are the one-dimensional distribution of magnetic energy density and electric field strength averaged over the $y$-$z$ plane. The solid and dashed lines correspond to the cases of $\rho_{\rm cl,S}/\rho_{\rm 0,S}=2$ and $10$. 
The left region of the shock front is the upstream region.
The magnetic field is amplified by the shock compression just behind the shock front and by the turbulence in more distant regions. 
The coherent length scale of the downstream turbulence is about $L_{\rm cl}$.
Interestingly, the saturation value of $\epsilon_{\rm B}=B_{\rm ,S}^2/4\pi\Gamma_{\rm up, S} \langle\rho_{\rm up, S}\rangle c^2$ is on the order of $10^{-4}$, which does not strongly depends on the density of clumps, $\rho_{\rm cl,S}/\rho_{\rm 0,S}$.
The value of $\epsilon_{\rm B}\sim 4\times 10^{-4}$ is a highly anticipated result in the GRB community \citep{Gruzinov1999, Huang2022}.\par


%
\begin{figure}[t]
    \centering
    \includegraphics[width=\linewidth]{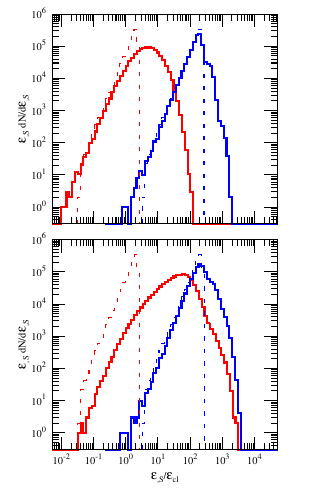}
    \caption{
    Energy spectra of test-particle simulations.
    The top and bottom panels are for $\rho_{\rm cl,S}/\rho_{\rm 0,S}=2$ and $10$, respectively.
    The red and blue solid histograms are the energy spectra for $R_{\rm up, U}/L_{\rm cl}=0.26$ and $=26$, respectively.
    The dotted histograms show the energy spectra without density fluctuations.
    }
    \label{fig:energy_spectrum_normalized}
\end{figure}
Figure~\ref{fig:energy_spectrum_normalized} is the energy spectra of particles in the test-particle simulation when they arrive at the downstream boundary. 
The particle energy is normalized by $\mathcal{E}_{\rm cl}=qB_{\rm up, U}L_{\rm cl}$, where $q$ is the particle charge.
The maximum energy in this system is limited by the distance between the shock front and the downstream boundary, $\mathcal{E}_{\rm max,S}\sim 10^{3}\mathcal{E}_{\rm cl}$.
The results with and without density variations are shown in the solid and dotted histograms, and the results for $R_{\rm up, U}/L_{\rm cl}=0.26$ and $26$ are shown in the red and blue histograms, respectively. 
For $R_{\rm up, U}/L_{\rm cl}=26$, some particles are accelerated but the peak of the energy spectrum does not change significantly compared to the case without density fluctuations. 
On the other hand, for $R_{\rm up, U}/L_{\rm cl}=0.26$, the position of the spectral peak significantly moves to the higher energy region. 
Comparing the top ($\rho_{\rm cl,S}/\rho_{\rm 0,S}=2$) and bottom ($\rho_{\rm cl,S}/\rho_{\rm 0,S}=10$) panels, the particles are accelerated more when the amplitude of density fluctuations is larger. \par

\begin{figure}[tp]
    \centering
    \includegraphics[width=\linewidth]{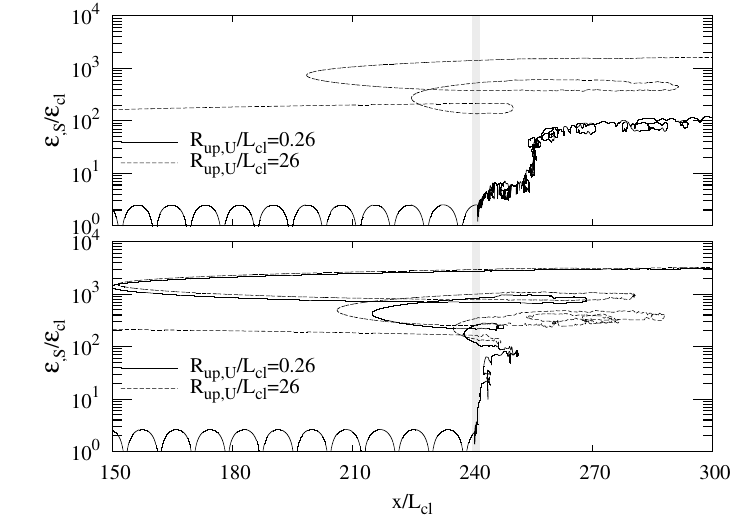}
    \caption{
    The representative particle trajectory for all the simulations. 
    The top and bottom panels are for $\rho_{\rm cl,S}/\rho_{\rm 0,S}=2$ and $10$, respectively. 
    The gray region represents the shock front.
    }
    \label{fig:trajectory}
\end{figure}
To understand the acceleration mechanism, we plot trajectories of the highest-energy particles for four runs in Figure~\ref{fig:trajectory}, where the horizontal and vertical axes show the particles' position in the mean shock normal direction and the particle energy. 
The gray region represents the shock front.
The solid and dotted lines are the trajectories for $R_{\rm up, U}/L_{\rm cl}=0.26$ and $=26$, respectively. 
For both cases of $\rho_{\rm cl,S}/\rho_{\rm 0,S}=2$ and $10$ with $R_{\rm up, U}/L_{\rm cl}=0.26$, the particles do not go back to the upstream region, but are accelerated in the downstream region in the range of $\mathcal{E}_{\rm ,S}\lesssim 100\mathcal{E}_{\rm cl}$.
This acceleration by the downstream turbulence is theoretically expected in early studies \citep{Ohira2013, Asano2015, Pohl2015, Yokoyama2020}.
The particle with the small gyroradius in the case of $\rho_{\rm cl,S}/\rho_{\rm 0,S}=10$ is efficiently accelerated just behind the shock front compared with one for $\rho_{\rm cl,S}/\rho_{\rm 0,S}=2$ (solid lines).
The fourth panel in Figure~\ref{fig:result_mhd_b_field} shows the electric field strength associated with the downstream turbulence, $\delta E_{\rm ,S}/E_{\rm up,S}$, where $\delta E_{\rm , S} = \langle|-(\delta \boldsymbol{u}_{\rm , S}/c)\times \boldsymbol{B}_{\rm , S}|\rangle_{yz}$, $E_{\rm up,S}=(u_{\rm up,S}/c)B_{\rm up,S}\sim B_{\rm up,S}$, and $\delta \boldsymbol{u}_{\rm , S}=\boldsymbol{u}_{\rm , S} - \langle \boldsymbol{u}_{\rm ,S}\rangle_{yz}$ is the fluid velocity associated with turbulence. 
$\langle Q \rangle_{yz}$ is a quantity $Q$ averaged over the $y$-$z$ plane. 
$u_{\rm up,S}$ is the upstream flow velocity.
The turbulent electric field for $\rho_{\rm cl,S}/\rho_{\rm 0,S}=10$ is stronger than one for $\rho_{\rm cl,S}/\rho_{\rm 0,S}=2$. 
The interaction with a denser clump generates stronger turbulence, resulting in faster magnetic field amplification, so that particles with the small gyroradius for $\rho_{\rm cl,S}/\rho_{\rm 0,S}=10$ are rapidly accelerated by the downstream turbulence.

For both cases of $\rho_{\rm cl,S}/\rho_{\rm 0,S}=2$ and $10$ with $R_{\rm up, U}/L_{\rm cl}=26$ and the case of $\rho_{\rm cl,S}/\rho_{\rm 0,S}=2$ with $R_{\rm up, U}/L_{\rm cl}=0.26$ in the range of $\mathcal{E}_{\rm ,S}\gtrsim 100~\mathcal{E}_{\rm cl}$, particles are accelerated by crossing the shock front many times. 
Our simulations show that the relativistic shock acceleration works if there are particles with a sufficiently large gyroradius in the vicinity of the shock front. 
This can be understood as follows. 
It takes the eddy turnover time, $T_{\rm eddy, S}\sim L_{\rm cl}/\delta u_{\rm ,S}$, for the shocked magnetic field to be disturbed.
Hence, the magnetic field is disturbed at the distance of $T_{\rm eddy, S}\times u_{\rm down, S}$ away from the shock front, but closer to the shock front than that, the magnetic field is almost uniform and perpendicular to the shock normal, where {\it $u_{\rm down, S}=u_{\rm 1, S}/r\approx c/r$} is the downstream flow velocity in the shock rest frame, and $r$ is the shock compression ratio.
To go back to the upstream from the downstream, the downstream gyroradius, $\mathcal{E}_{,\rm S}/qrB_{\rm up, S}$, has to be larger than $T_{\rm eddy, S}\times u_{\rm 2, S}$, which gives the injection energy for the shock acceleration,
\begin{equation}
\mathcal{E}_{\rm inj, S} \sim (c/\delta u_{\rm ,S})qB_{\rm up, S}L_{\rm cl}=(c/\delta u_{\rm ,S})\Gamma_{\rm up, S}\mathcal{E}_{\rm cl}.
\label{eq:injection_energy_for_shock_acc}
\end{equation}
Since our MHD simulations show $c/\delta u_{\rm ,S}\sim10$ in the downstream region, $\mathcal{E}_{\rm inj, S}\sim 50~\mathcal{E}_{\rm cl}$, which is consistent with results in our test-particle simulations.

\begin{figure}[tbp]
    \centering
    \includegraphics[width=\linewidth]{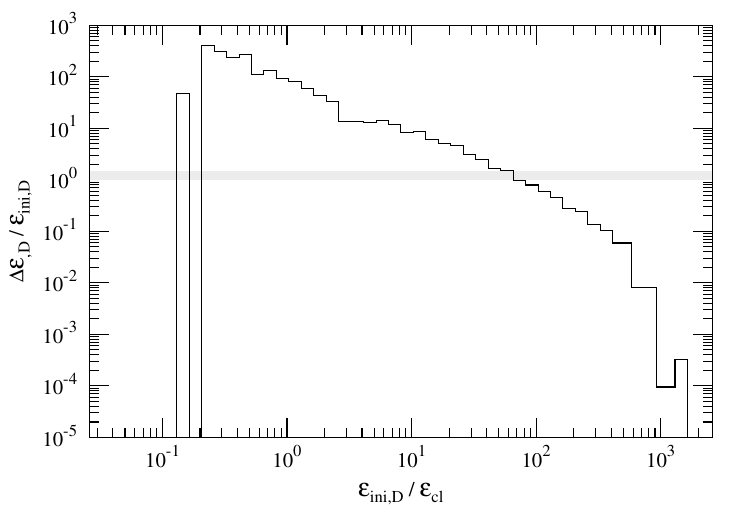}
    \caption{
    Mean downstream energy gain in the downstream rest frame for the top $10^3$ energetic particles in the case of $\rho_{\rm cl,S}/\rho_{\rm 0,S}=10$ and $R_{\rm up, U}/L_{\rm cl}=0.26$.
    The horizontal axis is the incident energy from upstream to downstream, $\mathcal{E}_{\rm ini,D}$.
    }
    \label{fig:energy_change_drf}
\end{figure}

For the case of $\rho_{\rm cl,S}/\rho_{\rm 0,S}=2$ with the initial small gyroradius, although some particles are accelerated to more than $\mathcal{E}_{\rm inj,S}$ in the far downstream region, they can not go back to the upstream region because they have already passed the diffusion length.
In contrast, for $\rho_{\rm cl,S}/\rho_{\rm 0,S}=10$ with the initial small gyroradius, the particle is quickly accelerated to $\mathcal{E}_{\rm inj,S}$ around the shock front, so that the particle can experience the relativistic shock acceleration.
To show that most particles are similarly accelerated, first, we extract parts of the trajectories of the top $10^3$ most energetic particles moving from upstream to downstream and back upstream again.
Then, we sample the incident energy from upstream to downstream, $\mathcal{E}_{\rm ini, D}$, and the downstream energy gain, $\Delta \mathcal{E}_{\rm ,D}$, in the downstream rest frame. 
For the shock acceleration without turbulent acceleration, $\Delta \mathcal{E}_{\rm ,D}=0$ and the mean energy gain during the one cycle is about $\mathcal{E}_{\rm ini,D}$. 
Figure~\ref{fig:energy_change_drf} shows $\Delta \mathcal{E}_{\rm ,D}/\mathcal{E}_{\rm ini,D}\gtrsim 1$ for $\mathcal{E}_{\rm ini,D}\lesssim 50~\mathcal{E}_{\rm cl}$, but $\Delta \mathcal{E}_{\rm ,D}/\mathcal{E}_{\rm ini,D}\lesssim 1$ for $\mathcal{E}_{\rm ini,D}\gtrsim 50~\mathcal{E}_{\rm cl}$. 
Therefore, for $\rho_{\rm cl,S}/\rho_{\rm 0,S}=10$ with the initial small gyroradius, most particles are accelerated first by the downstream turbulence and then by the relativistic shock. 
The injection energy to the shock acceleration is consistent with Equation~(\ref{eq:injection_energy_for_shock_acc}). 
This injection process to the shock acceleration has never been observed in previous studies. The detailed mechanism of the downstream acceleration will be addressed in future work.

\begin{figure}[tbp]
    \centering
    \includegraphics[width=\linewidth]{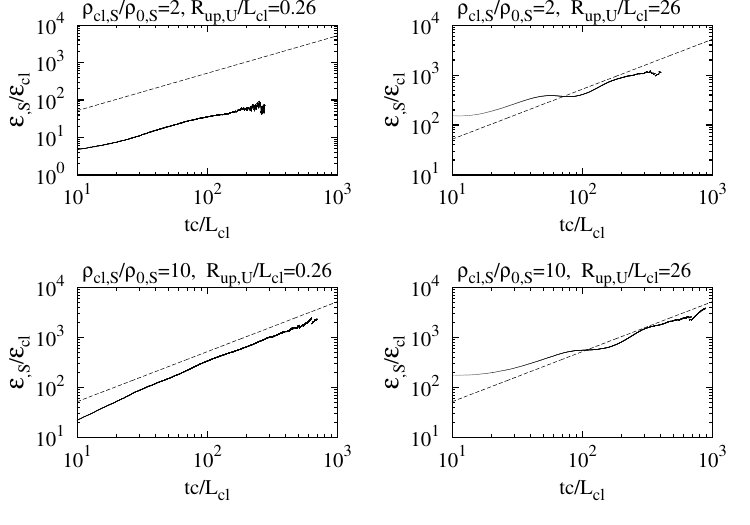}
    \caption{
    Mean time evolution of the particle energy for the four runs.
    $t=0$ is defined by the time when each particle crosses the shock front for the first time.
    The solid curves are the simulation results averaged for the top 1000 particles. The dashed lines are the theoretical line of the Bohm rate in the upstream magnetic field strength, where the initial energy was ignored.}
    \label{fig:energy_evolution}
\end{figure}

Finally, we show the mean evolution of the energy for the top 1000 energetic particles in Figure~\ref{fig:energy_evolution}, where the solid and dashed curves are the simulation results and theoretical curve of the Bohm rate in the upstream magnetic field strength, $(d\mathcal{E}/dt)_{\rm ,S}=qB_{\rm up,S}c$. 
Except for the top left panel, the simulation results are very close to the theoretical curve in the late phase when particles are accelerated by the relativistic shock. 
This is the first simulation that reproduces the ideal fast acceleration of CRs in the relativistic shock by calculating the magnetic field amplification and particle trajectory.
In the relativistic shock acceleration, particles double their energy for every round trip of a relativistic shock \citep{Achterberg2001}. The downstream residence time is negligible because the downstream magnetic field is strongly amplified in our simulation. In this case, the relativistic shock acceleration rate is given by the Bohm rate in the upstream magnetic field strength \citep{Kamijima2020}.

%


\section{Discussion and conclusion}
\label{sec:4}
The evolution of the magnetic field amplification and downstream turbulence depends on the resolution of MHD simulations \citep[e.g.][]{Ji2016}.
Our MHD simulations did not have enough resolution to converge all the results.
Nevertheless, we can theoretically explain our results concerning the injection to acceleration by relativistic shocks with Equation~(\ref{eq:injection_energy_for_shock_acc}).
Therefore, most results on particle acceleration are qualitatively robust although dependency of the clump density and initial gyroradius may change quantitatively.

In Figure \ref{fig:energy_evolution}, we showed that the ideal rapid acceleration works in the relativistic shock. 
If the downstream gyroradius is larger than the coherence length scale of magnetic field turbulence, $L_{\rm cl}$, the scattering efficiency is smaller than the Bohm rate in the downstream magnetic field strength. 
In this case, the downstream diffusion coefficient, $\kappa$, is given by 
\begin{equation}
\kappa \sim L_{\rm cl}c\left(\frac{r_{\rm down,S}}{L_{\rm cl}}\right)^2~,
\label{eq:diffusion_coefficient}
\end{equation}
where $r_{\rm down,S}$ is the downstream gyroradius. 
Then, the downstream residence time of accelerated particles is roughly given by $t_{\rm d,S} \sim \kappa / c^2$ \citep[e.g.][]{Sironi2013, Lemoine2019,Vanthieghem2020}.
On the other hand, the upstream residence time of accelerated particles is given by $t_{\rm u,S} \sim \Omega_{\rm up,S}^{-1}$ \citep{Achterberg2001}. 
If the upstream residence time is longer than the downstream residence time, the acceleration time scale is given by the upstream residence time, that is, the acceleration rate is given by the Bohm rate in the upstream magnetic field strength \citep{Kamijima2020}. 
From the condition, $t_{\rm u,S} > t_{\rm d,S}$, the energy range in which the ideal acceleration at the Bohm rate works is given by
\begin{eqnarray}
\frac{\mathcal{E}_{\rm, S}}{\mathcal{E}_{\rm cl}}~\lesssim~2000 ~\left(\frac{\bar{B}_{\rm down, S}}{20B_{\rm up,S}}\right)^{2}\left(\frac{\Gamma_{\rm up, S}}{5}\right)
\label{eq:Bohm_acceleration_energy}
\end{eqnarray}
where $\bar{B}_{\rm down, S}\sim20~B_{\rm up,S}$ is the average downstream magnetic field in our MHD simulation. 
If the particle energy is larger than the above range, the acceleration rate is slower than the Bohm rate in the upstream magnetic field. 

We showed in Figure \ref{fig:trajectory} that some particles are accelerated by the relativistic shock, but Figure \ref{fig:energy_spectrum_normalized} does not show a clear power-law spectrum. 
This is probably due to the limitation of the acceleration region. 
The maximum energy is at least limited by the finite size of the downstream region, $L_{\rm down}$, and given by the condition, $\kappa /c \sim L_{\rm down}$. 
Then, from Equation (\ref{eq:diffusion_coefficient}), the size-limited maximum energy is estimated to $\mathcal{E}_{\rm max,S}/\mathcal{E}_{\rm cl} \sim 800 ~ (L_{\rm down}/60L_{\rm cl})^{1/2}(\Gamma_{\rm up,S}/5)(\bar{B}_{\rm down,S}/20B_{\rm up,S})$.
We cannot expect a clear power-law spectrum above this energy, which is consistent with our results.
If the upstream density clumps have different larger length scales, higher energy particles would be efficiently confined in the downstream region. 
A clear power-law energy spectrum could then be observed in the test particle simulation. 
To confirm this point, we need to perform simulations with a larger simulation box.
In addition, we have to investigate dependences of physical parameters ($\sigma,~\beta,~\Gamma_{\rm up, S}$), and other realistic density structures \citep{Hu2022}.

Our simulations showed that particles with the energy of 0.26 $\mathcal{E}_{\rm cl}$ are accelerated in the downstream region, where $\mathcal{E}_{\rm cl} \sim 10^{12}~{\rm eV} \left(B_{\rm up,U}/1~{\rm G} \right) \left( L_{\rm cl}/10^{10}~{\rm cm} \right)$. 
The clump size of $10^{10}$ cm in the stellar wind is suggested by several studies \citep{Sironi2007, Moens2022, Chene2020}. Since the energy scale of $\mathcal{E}_{\rm cl}$ is much higher than the thermal energy of upstream particles, there is a huge gap between thermal particles and our simulation particles. The time scale of the downstream acceleration is an increasing function of the particle energy (see Figure \ref{fig:energy_evolution}). We naively expect a similar energy dependence even in a smaller energy range, but our simulations cannot investigate whether thermal particles are accelerated mainly by the same mechanism or not. On the other hand, it has been shown by particle-in-cell simulations that the Weibel-mediated collisionless shock can accelerate thermal particles to higher energies \citep{Spitkovsky2008, Sironi2013}. In addition, the Weibel turbulence would suppress particle streaming along the magnetic field line. Particle streaming makes the high-density region smoother, resulting in weaker downstream turbulence \citep{Tomita2022}. The Weibel turbulence, which cannot be resolved by MHD simulations, therefore plays an important role in bridging the gap between the thermal particles and our test-particle simulations, and in keeping the amplitude of density fluctuations large. 


In summary, using three-dimensional relativistic MHD simulations and test-particle simulations, we have shown the following.
\begin{enumerate}
  \setlength{\parskip}{0cm} 
  \setlength{\itemsep}{0cm} 
  \item Relativistic shocks propagating to inhomogeneous media amplify the downstream magnetic field to $\epsilon_{\rm B}\sim 4\times 10^{-4}$. 
  \item Particles are rapidly accelerated by the relativistic shock and downstream turbulence in this system.
  \item If the amplitude of the upstream inhomogeneity is sufficiently large, low energy particles are initially accelerated by the downstream turbulence and then accelerated by the relativistic shock.
  \item The injection energy to the shock acceleration in this system is given by Equation~(\ref{eq:injection_energy_for_shock_acc}).
\end{enumerate}

These results have long been awaited to explain emissions from relativistic shocks such as GRBs and active galactic nuclei. 
Calculating electromagnetic waves, neutrinos, and CRs from this system and comparing them with observations will help us to understand the environment surrounding the high-energy objects, leading to a better understanding of the origin of extra-galactic CRs and enigmatic high-energy objects. 
Moreover, this work can be applied to non-relativistic perpendicular shocks realized in supernova remnants of massive stars. 
The injection to the shock acceleration at the perpendicular shock would be enhanced by the upstream density inhomogeneity. 

\begin{acknowledgments}
We are grateful to the referee for valuable comments to improve the paper.
In addition, we are grateful to Yosuke Matsumoto for sharing the SRCANS+ code.
We also thank Sara Tomita and Shota Yokoyama for MHD simulations.
K. M. is supported by JST SPRING, Grant Number JPMJSP2108 and supported by International Graduate Program for Excellence in Earth-Space Science (IGPEES), The University of Tokyo.
Y. O. is supported by JSPS KAKENHI Grants No. JP21H04487, and No. JP24H01805.
Numerical computations were carried out on Cray XC50 at Center for Computational Astrophysics, National Astronomical Observatory of Japan.
\end{acknowledgments}

\bibliography{reference}
\bibliographystyle{aasjournal}

\end{document}